\newif\ifsubmode
\submodetrue

\ifsubmode
  \documentclass[12pt,preprint]{aastex}
  \received{}
  \revised{}
  \accepted{}
\else
  \documentclass[12pt,preprint]{emulateapj}

\fi

\newcommand{\etal}{et al.}
\newcommand{\acsB}{\hbox{$B_{435}$}}
\newcommand{\acsV}{\hbox{$V_{606}$}}
\newcommand{\acsi}{\hbox{$i_{775}$}}
\newcommand{\acsz}{\hbox{$z_{850}$}}
\newcommand{\longidrop}{CDFS~J033240.0$-$274815}
\newcommand{\idrop}{SiD2}

\shorttitle{$z \approx 6$ galaxies from GOODS}
\shortauthors{Dickinson et al.}

\begin{document}

\title{Color--selected galaxies at \mbox{\boldmath $z \approx 6$} 
in the Great Observatories Origins Deep Survey
\altaffilmark{1}}

\author{M.\ Dickinson\altaffilmark{2}, 
D.\ Stern\altaffilmark{3}, 
M.\ Giavalisco\altaffilmark{2},
H.\ C.\ Ferguson\altaffilmark{2},
Z.\ Tsvetanov\altaffilmark{4},
R.\ Chornock\altaffilmark{5}, 
S.\ Cristiani\altaffilmark{6},
S.\ Dawson\altaffilmark{5}, 
A.\ Dey\altaffilmark{7},
A.\ V.\ Filippenko\altaffilmark{5}, 
L.\ A.\ Moustakas\altaffilmark{2},
M.\ Nonino\altaffilmark{6},
C.\ Papovich\altaffilmark{8},
S.\ Ravindranath\altaffilmark{2},
A.\ Riess\altaffilmark{2}
P.\ Rosati\altaffilmark{9},
H.\ Spinrad\altaffilmark{5},
E.\ Vanzella\altaffilmark{9,10}
}

\altaffiltext{1}{Based on observations taken with the NASA/ESA Hubble
  Space Telescope, which is operated by AURA, Inc., under NASA 
  contract NAS5--26555; from the W.\ M.\ Keck Observatories; and 
  from the Very Large Telescope (VLT) at Cerro Paranal, Chile, 
  operated by the European Southern Observatory, under programs 
  170.A-0788 and 168.A-0485.}

\altaffiltext{2}{Space Telescope Science Institute (STScI), 
	3700 San Martin Dr., Baltimore, MD 21218.}
\altaffiltext{3}{Jet Propulsion Laboratory, California Institute of Technology,
	Mail Stop 169-506, Pasadena, CA 91109}
\altaffiltext{4}{Department of Physics and Astronomy, The Johns Hopkins
	University, 3400 N. Charles St., Baltimore, MD 21218--2686}
\altaffiltext{5}{Department of Astronomy, University of California, Berkeley, 
	Mail Code 3411, Berkeley, CA 94720}
\altaffiltext{6}{Istituto Nazionale di Astrofisica, Osservatorio Astronomico di
	Trieste, via G.B. Tiepolo 11, Trieste, I--34131, Italy}
\altaffiltext{7}{National Optical Astronomical Observatory, 950 North
	Cherry Avenue, Tucson, AZ 85719}
\altaffiltext{8}{Steward Observatory, University of Arizona, 933 Cherry Ave.,
	Tucson, AZ 85721--0065}
\altaffiltext{9}{European Southern Observatory, Karl--Schwarzechild--Strasse 2,
	D--85748 Garching bei M\"unchen, Germany}
\altaffiltext{10}{Dipartimento di Astronomia dell'Universit\`a di Padova, 
 	Vicolo dell'Osservatorio 2, I-35122 Padova, Italy}


\begin{abstract}

We report early results on galaxies at $z \sim 6$, selected from 
{\it Hubble Space Telescope} imaging for the Great Observatories Origins 
Deep Survey.  Spectroscopy of one object with the Advanced Camera for Surveys 
grism and from the Keck and VLT observatories a shows a strong continuum 
break and asymmetric line emission, identified as Ly$\alpha$ at $z = 5.83$.   
We detect only five spatially extended, $z \sim 6$ candidates with
signal--to--noise ratios $> 10$, two of which have spectroscopic 
confirmation.  This is many fewer than would be expected if galaxies 
at $z = 6$ had the same luminosity function as those at $z = 3$.
There are many fainter candidates, but we expect substantial contamination 
from foreground interlopers and spurious detections.  Our best estimates 
favor a $z = 6$ galaxy population with fainter luminosities, higher 
space density, and similar co--moving ultraviolet emissivity to that 
at $z = 3$, but this depends critically on counts at 
fluxes fainter than those reliably probed by the current data.

\end{abstract}
 


\section{Introduction
\label{section:intro}}

Broad band color selection, based on ultraviolet (UV) spectral breaks 
caused by neutral hydrogen, is an efficient technique for identifying 
galaxies at $z = 3$ to 4 with imaging from the ground and from the 
{\it Hubble Space Telescope} \citep{ste96, mad96}.  
At higher redshifts and 
relatively bright magnitudes, $i^\prime-z^\prime$ colors from the Sloan 
Digital Sky Survey have been used to identify QSOs out to $z = 6.4$ 
\citep{fan03}.  Some galaxies at $z > 5$ have also been found in this 
way, but the required deep imaging and spectroscopy is extremely 
challenging.   A Lyman break galaxy (LBG) with typical ($L^\ast$) 
UV luminosity at $z = 3$ ($M_{\rm 1700\AA} = -21.0$, 
Adelberger \& Steidel 2000) would have $m(z) = 26.0$ if moved, 
without evolution, to $z = 6$, and would be undetected in the 
$i$--band (hence, an ``$i$--dropout'').  At $z \gtrsim 6.5$, 
Ly$\alpha$ shifts through the $z$--band, and galaxies are 
lost to optical sight altogether.

One goal of the Great Observatories 
Origins Deep Survey (GOODS) is to find and study large numbers of 
galaxies at $3.5 < z < 6.5$.  Here, we report initial 
results from GOODS on galaxy candidates at $z \sim 6$, including 
spectroscopy for one galaxy, \longidrop\ (henceforth \idrop), 
with the ACS grism and the Keck and VLT observatories.  
We use AB magnitudes 
(AB~$\equiv 31.4 - 2.5\log\langle f_\nu / \mathrm{nJy} \rangle$), 
and assume a cosmology with 
$\Omega_{\rm tot}, \Omega_M, \Omega_\Lambda = 1.0, 0.3, 0.7$
and $H_0 = 70$~km~s$^{-1}$~Mpc$^{-1}$.

\section{Imaging, photometry, and color--selection
\label{section:imaging}}

The GOODS Treasury program covers areas around the Chandra Deep Field South
(CDF--S) and Hubble Deep Field North (HDF--N) with mosaics of ACS images.
The observations, data reduction, and catalogs are described in \citet{gia03a}.
Our present analysis uses 3--epoch co-added images for both fields, 
with 3, 1.5, 1.5 and 3 orbit depth in the F435W, F606W, F775W, and F850LP 
filters (henceforth \acsB, \acsV, \acsi, and \acsz).  After discarding 
regions near the image borders or without 4--band coverage, the survey 
solid angle is 316~arcmin$^2$.  We detect objects in the $\acsz$ images 
using SExtractor \citep{ber96}, and measure photometry through matched 
apertures in all bands.  Here, we use $\acsz$ ``total'' magnitudes 
(SExtractor MAG\_AUTO), and colors measured through isophotal apertures 
defined in the $\acsz$ image.   

We estimate the reddest colors expected for ordinary galaxies 
with spectral templates \citep{col80} integrated through the 
ACS bandpasses.  The redshifted colors of an elliptical galaxy 
peak\footnote{
The redshifted elliptical template has redder colors at 
$1.7 < z < 2.3$, but the UV spectrum of any galaxy at that 
redshift is unlikely to resemble that of an old elliptical 
at $z\approx 0$.
} 
at $\acsi - \acsz \approx 1.2$ for $z \approx 1.1$.
There is only one ``bright'' galaxy in the GOODS fields
with $\acsi - \acsz > 1.3$ ($\acsz = 23.9$; $\acsi - \acsz = 1.32$).
It is well detected at $\acsV$, bright in the near--infrared (IR), 
and certainly has $z \ll 6$.   Redder colors may be explained 
by dust obscuration, high metallicity, strong line emission in the 
$\acsz$--band, or intergalactic medium (IGM) absorption at high 
redshift.  For the range of intrinsic UV colors 
seen for LBGs at $z \approx 3$, $\acsi - \acsz > 1.3$ is crossed 
at $z = 5.5$ to 5.7.  Cool stellar dwarfs may also be this red.
\citet{fan03} Figure~2 shows that only a tiny minority of 
high-latitude stars have $i^\prime - z^\prime > 1.3$, and our 
ACS imaging provides a robust measure of stellarity for $\acsz < 26.2$.  

Our current ACS mosaics have very small misalignments between images 
from different observing epochs.  These can trigger over--rejection 
in the cores of point sources during cosmic ray removal in the \acsV\ 
and \acsi\ bands (only -- the \acsB\ and \acsz\ images are reduced 
differently).  There is virtually no photometric impact for extended 
sources \citep{gia03a}, but the $\acsi - \acsz$ colors of brighter stars 
can be biased redward, and we treat them with caution here.  

\begin{figure}
\epsscale{0.9}
\plotone{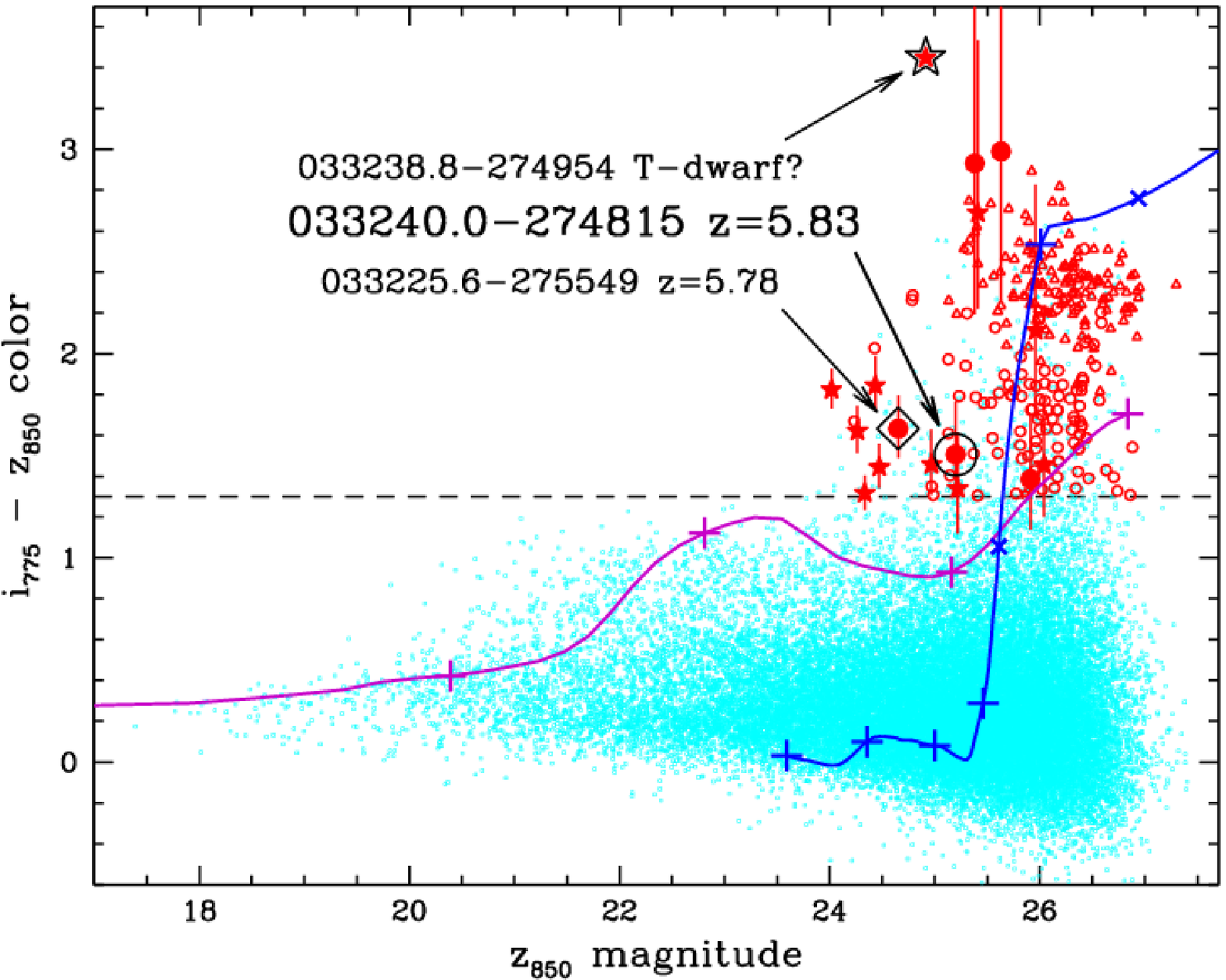}
\caption{Color--magnitude diagram for the GOODS CDF-S + HDF-N 
fields.  The dashed line shows our $i$--dropout color selection limit.
Light blue points are extended objects that do not meet the color and 
$S/N(\acsB,\acsV) < 2$ criteria.  Triangles mark $1\sigma$ lower color limits 
for objects undetected in $\acsi$.  Red points are $i$--dropout candidates.
Larger, filled points with error bars ($1\sigma$) are candidates with 
$S/N(\acsz) > 10$; stars mark point sources.  Two galaxies with spectroscopic 
redshifts are highlighted, as is a possible T--dwarf.  The mauve curve 
shows the color--magnitude track for an unevolving, $L^\ast$ elliptical 
galaxy; vertical crosses mark $z = 0.5$, 1, 1.5 and 2.0.  The blue curve 
shows the track for an unevolving $L^\ast_{z=3}$ LBG with average UV 
colors;  vertical crosses mark $z = 2$, 3, 4, 5, and 6, while tilted 
crosses mark $z = 5.5$ and 6.5.
}
\label{fig:cm_acs}
\end{figure}

We are interested in objects near our detection limits.  The 
signal--to--noise ratio ($S/N$) of a measurement depends on the source 
flux and size, and on the exposure time, which varies with position 
in our mosaics.  The significance of a source is best 
estimated not from its magnitude, but from $S/N(\acsz)$ in the detection 
aperture.  Our photometric errors are computed from noise maps which 
account for inter-pixel correlations.  To verify their reliability, 
we added artificial objects to the $\acsz$ images (only) and detected 
them with SExtractor.  Background--subtracted counts ($S_i$) were 
measured through matched apertures for the other bands, 
and compared to the predicted uncertainties ($\sigma_i$) from the noise 
maps.  The distribution of $S_i/\sigma_i$ is nearly Gaussian with 
mode~$\approx 0$ and RMS~$ \approx 1$, showing that our error estimates 
are reliable, except for a positive tail due to blending 
with other objects.  Because of this tail, 14\% of $z \sim 6$ galaxies 
would have $S/N > 2$ in the \acsB\ or \acsV\ bands.\footnote{On average, 
this limit corresponds to \acsB\ or $\acsV > 29.1$.}
We consider this an acceptable loss rate, and adopt $\acsi - \acsz > 1.3$ 
and $S/N(\acsB,\acsV) < 2$ as our $i$--dropout criteria 
(Figure~\ref{fig:cm_acs}).  

\section{Spectroscopy}

\begin{figure}
\epsscale{0.75}
\plotone{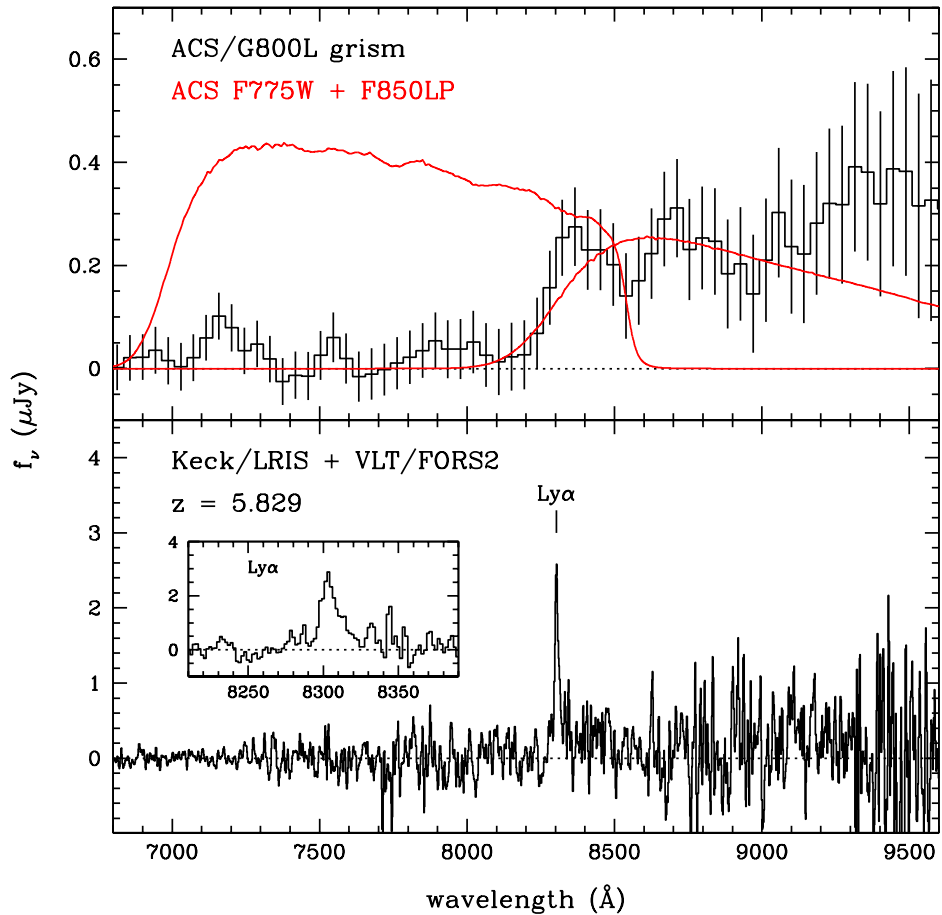}
\caption{{\it (a) Top}:  ACS grism spectrum of \idrop.  Pixels in 
the spectrum are correlated by the data reduction process, and thus 
have smaller scatter than suggested by the $1\sigma$ error bars.
Curves show unnormalized bandpass functions for the $\acsi$ and 
$\acsz$ filters.  
{\it (b) Bottom:}  Keck + VLT spectrum of \idrop, slightly smoothed.
The inset panel shows a magnified (unsmoothed) view of the Ly$\alpha$
emission line.
}
\label{fig:spectra}
\end{figure}

Supernovae found in GOODS ACS imaging are studied in 
a target--of--opportunity program which, in some cases, obtains 
low--resolution, slitless spectra with the ACS G800L grism.  One 
grism observation (SN2002FW, \citet{rie03}) included \idrop, which 
we had noted as an $i$--dropout candidate.  The data were obtained 
on UT 2002 October 01, with an exposure time of 18840s, and were 
reduced with the {\tt calacs} pipeline and the {\tt aXe} extraction 
software.  The G800L spectrum (Figure~\ref{fig:spectra}$a$) shows 
a flat continuum with a sharp break at $\lambda \approx 8300$\AA, 
explaining the red $\acsi - \acsz$ color.   

We observed \idrop\ with the Low Resolution Imaging Spectrometer 
(LRIS; \citet{oke95}) on the Keck~II telescope on UT 2002 October 09
in poor weather conditions, but detected line emission at 8303\AA.  
Deeper observations (7.8 ks of integration) with the 400 lines mm$^{-1}$ 
grating ($\lambda_{\rm blaze} = 8500$ \AA; $R \approx 1000$) were 
obtained on UT 2002 November 08.  Observations (12~ks) with the 
Focal Reducer/Low Dispersion Spectrograph (FORS2) on the Yepun 
telescope (VLT~4) were obtained on UT 2002 December 08 with the 300I 
grism ($R \approx 860$).  
We reduced the data with IRAF following standard procedures, and 
combined the LRIS and FORS2 data with appropriate weighting.  
The final spectrum (Figure~\ref{fig:spectra}$b$) shows Ly$\alpha$ emission 
at $z = 5.829$ with flux $1.6 \times 10^{-17}~{\rm erg}~{\rm cm}^{-2}~{\rm s}^{-1}$.  
The line shows the blue cut--off characteristic of high--redshift 
Ly$\alpha$ emitters, and the Ly$\alpha$ forest continuum break is 
clearly evident.  The emission line is not obviously detected in the 
grism spectrum.  The ACS exposure time calculator predicts a line 
detection with $S/N \approx 15$ for a point source.  Extended line 
emission superimposed on the galaxy continuum, however, is evidently 
invisible.

We also obtained a spectrum of the only spatially extended HDF--N candidate 
with $S/N(\acsz) > 10$ (J123619.9+620934) with Keck/LRIS on UT 2003 May 01, 
but did not successfully measure a redshift.

\section{Other candidate $z \approx 6$ objects
\label{section:candidates}}

\begin{figure}
\epsscale{0.9}
\plotone{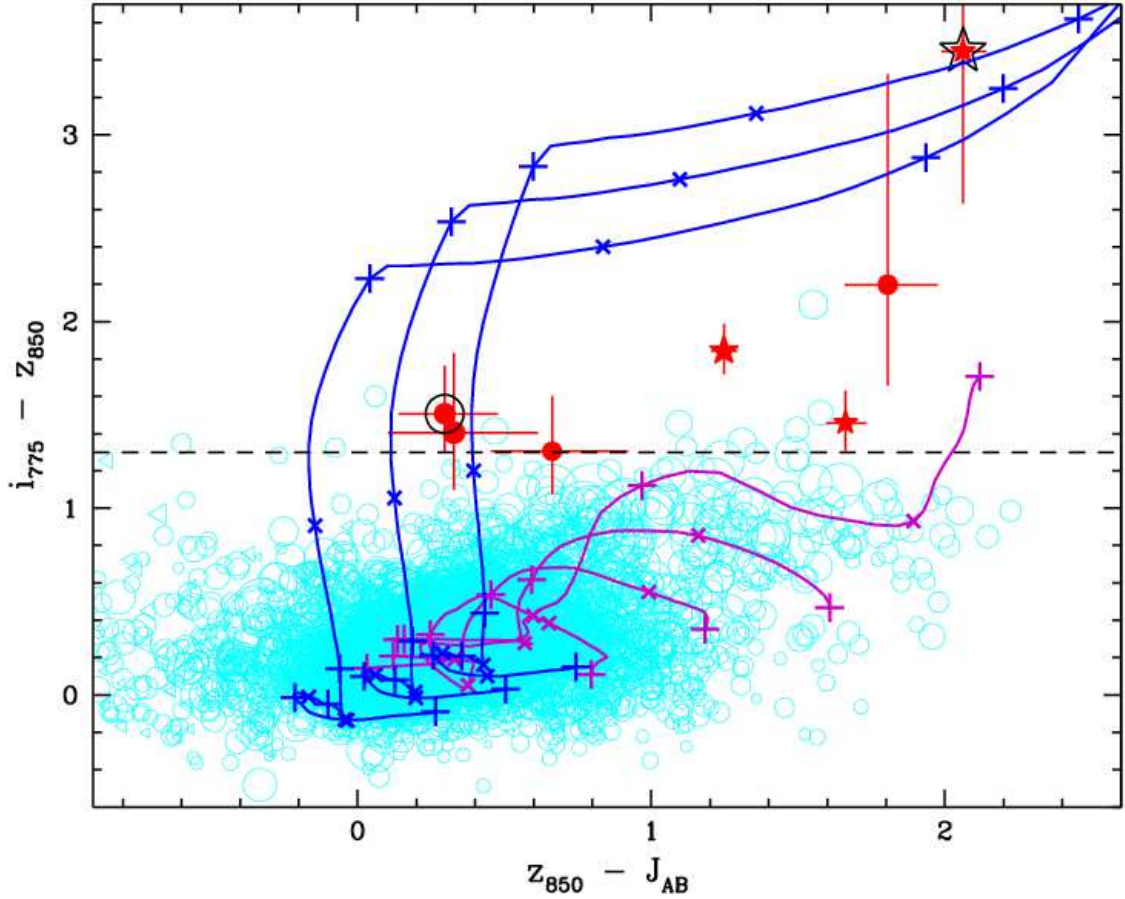}
\caption{Optical--infrared 2--color diagram for the portion
of the CDF--S field with deep ISAAC near--IR data.  Points 
show objects with $S/N(z) > 7$, with sizes proportional to their 
$J$ magnitudes.  Filled circles and stars with error bars 
($1\sigma$) are objects (extended and unresolved) that meet our 
$i$--dropout criteria.  The point for \idrop\ is circled.  
Open circles show extended objects which do not meet these criteria.  
Mauve tracks show modeled colors of ordinary, low--redshift galaxies, 
redshifted over $0 < z < 2$.  Blue tracks show expected colors 
of LBGs at $2 < z < 7$ ($z = 6$ at the ``bend''), spanning the range 
of UV spectral slopes seen in LBGs at $z \approx 3$.  Crosses on 
the tracks mark the same redshifts as in Figure~1.
}
\label{fig:cc_isaac}
\end{figure}

There are 16 objects with $S/N(\acsz) > 10$ that meet our selection 
criteria.  11 are point sources (3.4\% of the stellar objects with 
$24 < \acsz < 26.2$;  $0\farcs12 < {\rm FWHM} < 0\farcs16$ versus 
$0\farcs18$--$0\farcs7$ for the other $S/N > 10$ candidates), whose 
$\acsi - \acsz$ colors are suspect (\S\ref{section:imaging}).   
The GOODS project has gathered deep near--IR imaging with the VLT ISAAC 
camera covering $\sim 30\%$ of our CDF--S field \citep{gia03a}, including 
all three CDF--S stellar $i$--dropout candidates.  Their $\acsz - J$ 
colors are too red compared to expectations for high--redshift objects 
(Figure ~\ref{fig:cc_isaac}).  Although some of the 8 HDF--N point sources 
might conceivably be high redshift objects, we believe that they
are probably stars and will not consider them further here.  This leaves 
five extended $z \sim 6$ candidates, or 0.016~arcmin$^2$ 
(Table~\ref{tab:candidates}).

\citet[SBM]{sta03} used the public release v0.5 GOODS ACS CDF--S data 
to identify nine $i$--dropout candidates. Three are in our sample, 
and two have been confirmed spectroscopically \citep[this paper]{bun03}.
SBM\#5 (J033238.80$-$274953.6) is unresolved, with the reddest $\acsi - \acsz$ 
color ($> 2.7$ at 2$\sigma$) of any GOODS object.
Its exceptionally blue near--IR colors ($J-H = -0.3$, $H-K = -0.5$, AB) 
suggest that it may be a T--dwarf (see also SBM).  Another point source, 
SBM\#4, was observed in the GOODS spectroscopic program and is a cool 
star (approximately L0V). SBM\#2, 4, 8 and 9 have $S/N > 2$ in \acsV\ 
and/or \acsB, are fairly bright in near--IR images, and are thus unlikely 
to be at $z \sim 6$.  SBM\#6 falls outside the area analyzed here, where 
the $Viz$ data are shallow and there are no \acsB\ data.  In summary, 
three (perhaps four) of the nine SBM objects are good $z\sim 6$ candidates.

Data artifacts (space junk trails, reflection ghosts, diffraction spikes, 
residual cosmic rays) can produce spurious $\acsz$ detections without 
shorter--wavelength counterparts which mimic $i$--dropouts.  
We have removed most of these by visual inspection.
This is generally easy at $S/N > 10$, but this corresponds to 
$\langle \acsz \rangle \approx 25.3$, which is fairly bright for galaxies 
at $z \approx 6$.  Our catalogs push deeper; {\it post--facto}, we truncate 
them at $S/N \geq 5$ and reject sources that are too small or sharp to 
be real.  Even after careful inspection, however, some spurious sources 
probably remain.  As one check, we masked areas with objects, inverted 
the images, and detected ``negative sources.'' We find 57 which qualify 
as $i$--dropouts.  All have $-S/N < 8$, and $\sim 75\%$ have $5 < -S/N < 6$.    

The vast majority of real, faint galaxies have $\acsi-\acsz < 1.3$ 
and $z \ll 6$, but measurement errors may scatter a small fraction 
to redder colors.  We estimate this contamination using brighter objects 
($S/N(\acsz) > 20$).  We randomly assign their colors to fainter objects, 
then perturb the simulated fluxes using the error distributions quantified 
in \S\ref{section:imaging}.  Only $\sim 2$ foreground interlopers 
with $S/N > 10$ would (barely) meet the $i$--dropout criteria, 
while $\sim 49$ objects with $5 < S/N < 10$ could do so.

Together, these contaminants represent $< 0.7\%$ of the $\sim 16000$
GOODS sources with $5 < S/N(\acsz) < 10$, but may contribute 
$\sim$45\% of the faint $z\sim 6$ candidates.  After subtracting 
the expected contamination, we estimate that there are $\sim 145$ 
candidates with $S/N > 5$ (0.46~arcmin$^{-2}$), $> 50\%$ of which 
have $5 < S/N < 6$.

There are four extended candidates with $S/N > 7$ in the portion of 
the CDF--S with deep ISAAC imaging (Figure ~\ref{fig:cc_isaac}).  
One has red $\acsz - J$ and bright IR magnitudes, and is unlikely 
to be at $z \approx 6$.  Two or perhaps three candidates, including 
\idrop, are very faint in the near--IR ($24.7 < J_{AB} < 24.9$), 
with colors expected for galaxies at $5.5 < z < 6$.

\section{Discussion}

The $\acsi - \acsz$ color limit sets a lower redshift bound 
for $i$--dropouts, while IGM suppression in \acsz\ makes the upper 
bound, and hence the sampling volume, a strong function of luminosity.
We use simulations \citep{gia03b} to predict number counts of candidates,
including photometric biases.  We generated artificial galaxies with 
a mixture of disk and bulge surface brightness profiles, ellipticities, 
and orientations.  Their sizes were drawn from a log--normal distribution
tuned to reproduce measurements at $z \approx 3$--5 \citep{fer03}.
Their spectra have a distribution of UV spectral slopes that matches 
the observed colors of LBGs at $z \approx 3$ \citep{ade00}.  We distributed 
the galaxies in redshift, modulated their spectra by IGM opacity 
\citep{mad95}, convolved them with ACS point spread functions, added 
them to the GOODS images at various magnitudes, and recovered them 
with SExtractor.    

\begin{figure}
\epsscale{0.9}
\plotone{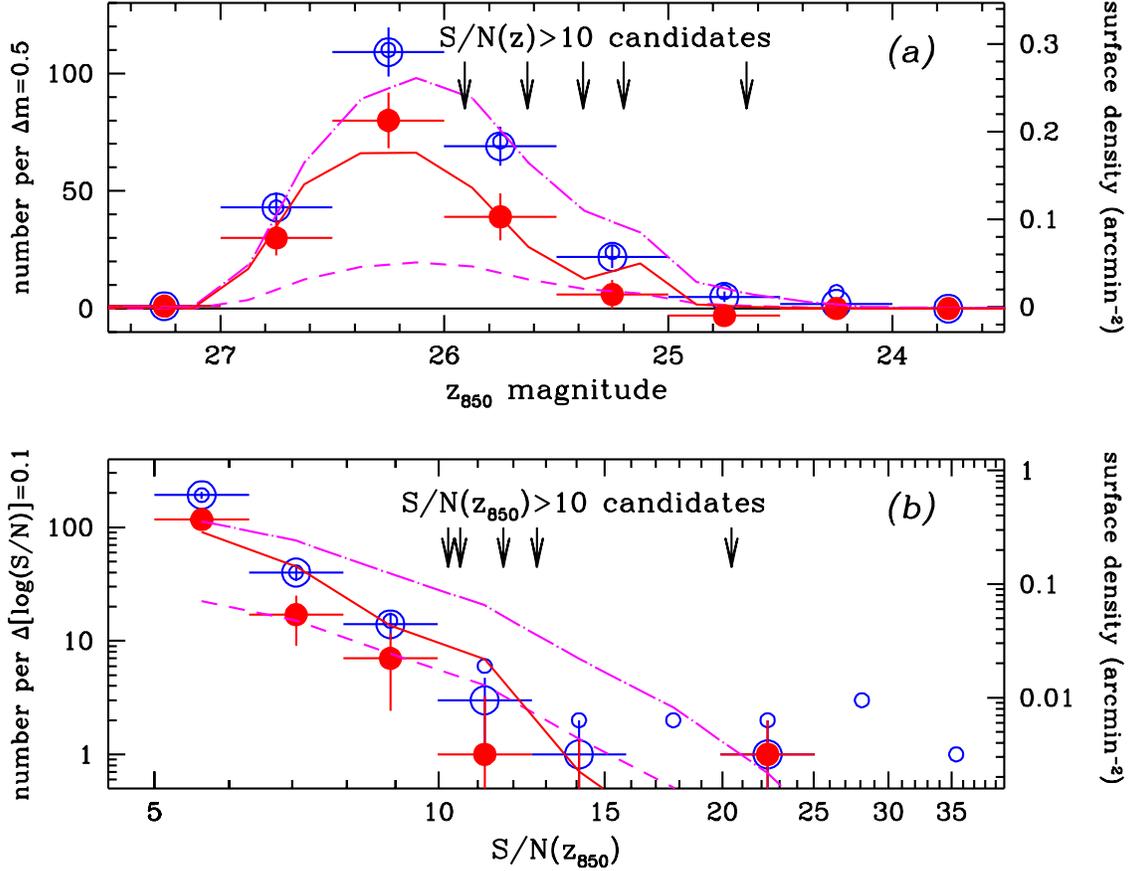}
\caption{
Number counts of $i$--dropout candidates vs.\ $\acsz$ magnitude 
({\it a}) and vs.\ $S/N$ the $\acsz$ detection image ({\it b}).
Small and large open circles show ``raw'' counts with and without 
point sources.  Filled points are counts after statistical correction 
for spurious objects.  Vertical error bars show $\sqrt N$ 
counting statistics.  Arrows indicate locations for the five most 
secure candidates.  Lines show predicted counts from simulations 
with various assumptions about the galaxy luminosity function.  
The dot--dashed line uses the $z=3$ LBG luminosity function, while 
the short--dashed line uses the same $L^\ast$ but reduces $\phi^\ast$ 
by a factor of 5.   The solid line shows the best--fit Schechter 
function to the corrected $N(\acsz)$ points, with 
$L^\ast = 0.4 L^\ast_{z=3}$ and $\phi^\ast = 3.8 \phi^\ast_{z=3}$.
}
\label{fig:s2ncount}
\end{figure}

Figure~\ref{fig:s2ncount} compares the number of $i$--dropout candidates 
to simulations for various assumptions about the UV luminosity function (LF),
which we model as a Schechter (1976) distribution with a faint--end slope 
fixed to $\alpha = -1.6$, as measured for LBGs at $z = 3$ \citep{ade00}.  
The number of {\it bright} galaxies is smaller at $z \approx 6$ than at 
$z = 3$ (see also SBM; Bremer \& Lehnert 2003 find a similar result from 
ground--based imaging for LBGs at $z \approx 5.3$).
The $z = 3$ LF is excluded with a high degree 
of confidence ($P < 10^{-8}$).  It predicts 30 galaxies with $S/N(\acsz) > 10$ 
vs.\ 5 observed, and 26 with $\acsz < 25$ vs.\ $\leq 7$ observed.\footnote{Out 
of 7 candidates with $\acsz < 25$, only one has $S/N(\acsz) > 10$.  The others 
may be real, but contamination may be substantial.}  A change in the number 
of bright objects does not require comparable evolution in the total luminosity 
density of the population; the number of bright sources is exponentially 
sensitive to the value of $L^\ast$.   Schechter functions fit to 
the counts in Fig.~\ref{fig:s2ncount}$a$ favor fainter $L^\ast$ and higher 
$\phi^\ast$ compared to their values at $z = 3$.  Integrating acceptable 
fits for $M_{\rm 1700\AA} < -19.4$ ($\approx 0.2 L^\ast_{z=3}$), the UV 
emissivity is similar to that at $z = 3$ 
($\rho(L_{z=6}) / \rho(L_{z=3}) = 0.77^{+0.29}_{-0.23}$, 95.4\% confidence).
However, these fitted $L^\ast$ values, and hence most of the inferred luminosity 
density, are at $\acsz > 26.4$, where the current data are most uncertain.
The LF fit is strongly driven by objects with $5 < S/N < 6.3$ 
(Fig.~\ref{fig:s2ncount}$b$).  A model with $L^\ast_{z=6} = L^\ast_{z=3}$ 
and $5\times$ smaller $\phi^\ast$ (and $\rho_L$) is consistent with the 
data at brighter magnitudes and higher $S/N$ ratios, but drastically 
underpredicts the counts at low $S/N$ and $\acsz > 25.5$.  Fits excluding 
the lowest $S/N$ bin leave the total $\rho_L$ essentially unconstrained.  
A robust determination of the $z = 6$ LF and total emissivity requires data 
significantly deeper than those used here.

Two other studies have analyzed $i$--dropouts from somewhat deeper 
ACS images.  \citet{yan03} found 2.3 candidates/arcmin$^2$ with 
$S/N(\acsz) > 7.2$ in an ACS field with exposure time 1.5$\times$ 
longer in \acsz\ than the 3--epoch GOODS data, but $4.9\times$ longer 
in \acsi, thus providing more robust color discrimination
against interlopers.   Their density is $10\times$ larger than ours 
to the same $S/N$ threshold.
They estimate their 
catalogs are 100\% complete for $\acsz \leq 28.0$, whereas ours 
are only 50\% complete for point sources at $\acsz = 26.7$ \citep{gia03a}.
Yan et al.\ may have underestimated their source fluxes or spurious 
detection rate, but it is notable that they also find very few 
bright candidates (none with $\acsz < 26.8$).  \citet{bou03} identified 
0.5 candidates/arcmin$^2$ with $\acsz < 27.3$ from imaging 
(5--20 orbits in \acsz) covering 46~arcmin$^2$.  They also find 
few bright candidates (only one with $\acsz < 25.5$), and estimate 
$\rho(L_{z=6}) / \rho(L_{z=3}) = 0.6\pm0.2$.

In summary, we have identified five spatially extended, 
high--$S/N$ candidates for galaxies at $z \sim 6$ in early 
GOODS ACS imaging.  Two have confirmed redshifts $z \approx 5.8$.
There are many fainter candidates, but we estimate that $\sim 45\%$ 
may be spurious detections or foreground interlopers.  The number 
of robust candidates is smaller than is predicted if the LF 
were the same as that at $z\approx 3$.  
Our best estimates find fainter $L^\ast$, larger $\phi^\ast$,
and moderately smaller $\rho_L$ compared to $z=3$, but this 
strongly depends on the number of objects at $\acsz > 26$, 
which is as yet poorly measured.  Constant $L^\ast$ with smaller
$\phi^\ast$ and $\rho_L$ are consistent with the bright counts 
but greatly underpredict the number of faint sources.  The measurements 
do not require (nor robustly exclude) a dramatic change in $\rho_L$
from $z \sim 6$ to 3, especially if $L^\ast$ is evolving with redshift.
\citet{gia03b} find only a modest change ($-30\%\pm10\%$) in the luminosity 
density from $z = 3$ to $z \approx 5$ where the GOODS LBG sample is 
much better characterized.  Our best estimates are consistent with an 
extrapolation of those results to $z = 6$, but deeper data are needed 
for a robust measurement.  The final GOODS images will be deeper, 
with fewer contaminating artifacts.  This, together with much deeper 
data (e.g., the forthcoming ACS Ultradeep Field), will provide better 
constraints on the galaxy population at these highest optically--accessible 
redshifts.

\acknowledgements

We thank the members of the GOODS team, and the staff at STScI, 
ESO and the Keck Observatory, who made this project possible.  Support 
was provided by NASA through grant GO09583.01-96A from STScI, which 
is operated by AURA under NASA contract NAS5-26555.  Work by LM and DS 
was supported by NASA through the {\it SIRTF} Legacy Science Program, 
through contract number 1224666, issued by JPL, California Institute 
of Technology, under NASA contract 1407.

\clearpage



\clearpage

\clearpage

\begin{deluxetable}{crrrcccl}
\tablewidth{0pc}
\tabletypesize{\scriptsize}
\tablecaption{$z \sim 6$ galaxy candidates, ordered by $S/N(\acsz)$\tablenotemark{a}
\label{tab:candidates}}
\tablehead{
\colhead{ID} &
\multicolumn{2}{c}{RA (J2000) Dec} &
\colhead{$S/N(\acsz)$} &
\colhead{$m(\acsz)$} &
\colhead{$\acsi - \acsz$\tablenotemark{b}} &
\colhead{FWHM(\acsz)} &
\colhead{Notes} \\
  & & & & & & \colhead{arcsec} & 
}
\startdata

SiD001 & 03:32:25.60 & 27:55:48.6 & 20.45 & $24.65 \pm 0.06$ & $1.63 \pm 0.15$ &  0.18 &
	SBM\#3 $z=5.78$ \citep{bun03}\\
SiD002 & 03:32:40.02 & 27:48:15.0 & 12.72 & $25.20 \pm 0.12$ & $1.51 \pm 0.23$ &  0.19 &
	SBM\#1 $z=5.83$ (this paper); ISAAC\\
SiD003 & 03:32:19.07 & 27:54:21.9 & 11.72 & $25.91 \pm 0.13$ & $1.39 \pm 0.28$ &  0.27 &
	SBM\#7; faint IR (SOFI)\\
NiD001 & 12:36:19.90 & 62:09:34.2 & 10.55 & $25.63 \pm 0.13$ & $>2.29$ &  0.19 & \\ 
SiD004 & 03:32:33.20 & 27:39:49.2 & 10.24 & $25.38 \pm 0.13$ & $>2.21$ &  0.70 &
	faint IR (SOFI)\\
\tableline
SiD005 & 03:32:39.03 & 27:52:23.1 &  9.44 & $25.29 \pm 0.17$ & $1.94 \pm 0.44$ &  0.28 &  \\ 
SiD006 & 03:32:45.23 & 27:49:09.9 &  9.05 & $25.91 \pm 0.22$ & $1.40 \pm 0.31$ &  0.65 &  \\ 
NiD002 & 12:37:28.62 & 62:20:39.1 &  8.79 & $24.43 \pm 0.11$ & $>2.01$ &  0.77 &  \\ 
SiD007 & 03:32:42.94 & 27:52:00.7 &  8.75 & $24.97 \pm 0.18$ & $1.35 \pm 0.30$ &  0.72 &  \\ 
NiD003 & 12:37:35.90 & 62:20:43.4 &  8.67 & $24.79 \pm 0.17$ & $>1.95$ &  0.93 &  \\ 
SiD008 & 03:32:13.06 & 27:49:00.7 &  8.63 & $25.78 \pm 0.16$ & $1.32 \pm 0.31$ &  0.34 &  \\ 
SiD009 & 03:32:41.36 & 27:50:04.7 &  8.58 & $26.19 \pm 0.21$ & $>2.07$ &  0.14 &  \\ 
SiD010 & 03:32:26.25 & 27:48:30.3 &  8.50 & $25.41 \pm 0.19$ & $1.31 \pm 0.26$ &  0.23 & ISAAC \\ 
NiD004 & 12:36:42.15 & 62:09:02.0 &  8.20 & $25.55 \pm 0.21$ & $1.59 \pm 0.52$ &  0.75 &  \\ 
NiD005 & 12:35:59.01 & 62:12:45.6 &  8.15 & $25.81 \pm 0.21$ & $1.55 \pm 0.36$ &  0.36 &  \\ 
NiD006 & 12:36:18.54 & 62:10:41.9 &  8.13 & $26.18 \pm 0.17$ & $1.92 \pm 0.52$ &  0.28 &  \\ 
NiD007 & 12:37:52.57 & 62:17:00.7 &  8.08 & $25.93 \pm 0.20$ & $>2.14$ &  0.21 &  \\ 
SiD011 & 03:32:37.63 & 27:50:22.4 &  8.07 & $25.15 \pm 0.21$ & $1.40 \pm 0.35$ &  0.88 & ISAAC \\ 
SiD012 & 03:32:23.84 & 27:55:11.5 &  8.02 & $26.23 \pm 0.18$ & $>1.98$ &  0.23 &  \\ 
NiD008 & 12:36:43.53 & 62:10:04.1 &  7.89 & $26.27 \pm 0.18$ & $>1.91$ &  0.19 &  \\ 
NiD009 & 12:37:12.43 & 62:18:28.4 &  7.65 & $25.94 \pm 0.23$ & $>2.04$ &  0.32 &  \\ 
NiD010 & 12:36:48.08 & 62:10:12.6 &  7.60 & $25.34 \pm 0.22$ & $>1.80$ &  0.90 &  \\ 
SiD013 & 03:32:34.69 & 27:50:22.8 &  7.57 & $25.31 \pm 0.23$ & $>1.94$ &  0.87 & ISAAC (bright: $J_{AB} = 23.1$, $K_{AB} = 21.6$) \\ 
NiD011 & 12:37:15.75 & 62:22:32.5 &  7.43 & $26.18 \pm 0.25$ & $>1.70$ &  0.13 &  \\ 
NiD012 & 12:36:48.71 & 62:12:17.1 &  7.41 & $25.37 \pm 0.16$ & $1.51 \pm 0.34$ &  0.57 &  \\ 
NiD013 & 12:36:13.04 & 62:10:43.6 &  7.34 & $25.89 \pm 0.22$ & $>1.89$ &  0.58 &  \\ 
NiD014 & 12:37:22.51 & 62:18:39.7 &  7.26 & $24.23 \pm 0.10$ & $1.67 \pm 0.50$ &  0.68 &  \\ 
NiD016 & 12:35:49.72 & 62:13:29.2 &  7.21 & $25.73 \pm 0.22$ & $>1.47$ &  0.61 &  \\ 
NiD015 & 12:35:50.89 & 62:11:58.8 &  7.21 & $26.02 \pm 0.23$ & $1.70 \pm 0.52$ &  0.32 &  \\ 
SiD014 & 03:32:52.22 & 27:48:04.8 &  7.20 & $26.42 \pm 0.27$ & $>1.81$ &  0.25 &  \\ 
NiD017 & 12:35:52.33 & 62:12:08.7 &  7.11 & $25.23 \pm 0.20$ & $1.79 \pm 0.55$ &  0.69 &  \\ 
SiD015 & 03:32:54.11 & 27:49:16.0 &  7.06 & $25.46 \pm 0.22$ & $>1.59$ &  0.47 &  \\ 
SiD016 & 03:32:11.93 & 27:41:57.1 &  7.04 & $26.38 \pm 0.26$ & $1.60 \pm 0.50$ &  0.27 &  \\ 
SiD017 & 03:32:33.16 & 27:41:17.1 &  6.95 & $25.54 \pm 0.23$ & $1.76 \pm 0.52$ &  0.64 &  \\ 
SiD018 & 03:32:54.06 & 27:51:12.0 &  6.91 & $26.37 \pm 0.21$ & $1.54 \pm 0.44$ &  0.19 &  \\ 
SiD019 & 03:32:36.34 & 27:43:15.6 &  6.82 & $26.18 \pm 0.18$ & $>1.78$ &  0.49 &  \\ 
NiD018 & 12:36:15.36 & 62:14:56.4 &  6.82 & $25.28 \pm 0.15$ & $>1.76$ &  0.18 &  \\ 
SiD020 & 03:32:22.27 & 27:52:57.2 &  6.80 & $25.32 \pm 0.25$ & $>2.03$ &  0.68 &  \\ 
SiD021 & 03:32:25.15 & 27:48:17.1 &  6.79 & $24.79 \pm 0.14$ & $>1.82$ &  0.58 & ISAAC \\ 
NiD019 & 12:37:08.89 & 62:19:19.1 &  6.78 & $26.79 \pm 0.29$ & $>1.73$ &  0.15 &  \\ 
NiD020 & 12:37:35.94 & 62:14:22.4 &  6.78 & $26.16 \pm 0.26$ & $>1.65$ &  0.55 &  \\ 
SiD022 & 03:32:29.84 & 27:52:33.2 &  6.77 & $26.03 \pm 0.22$ & $>1.72$ &  0.24 &  \\ 
SiD023 & 03:32:46.05 & 27:49:29.7 &  6.75 & $25.61 \pm 0.21$ & $>1.78$ &  0.32 &  \\ 
NiD022 & 12:37:12.94 & 62:18:05.6 &  6.75 & $25.71 \pm 0.23$ & $>1.96$ &  0.50 &  \\ 
NiD021 & 12:36:08.21 & 62:09:10.8 &  6.75 & $26.58 \pm 0.21$ & $>1.84$ &  0.13 &  \\ 
NiD023 & 12:37:25.35 & 62:18:45.6 &  6.70 & $25.83 \pm 0.21$ & $>1.71$ &  0.41 &  \\ 
NiD024 & 12:37:42.85 & 62:19:41.8 &  6.64 & $26.37 \pm 0.25$ & $>1.75$ &  0.52 &  \\ 
NiD025 & 12:36:35.63 & 62:09:35.8 &  6.62 & $26.81 \pm 0.22$ & $>1.46$ &  0.30 &  \\ 
SiD024 & 03:32:32.46 & 27:40:02.0 &  6.56 & $26.40 \pm 0.21$ & $>1.72$ &  0.30 &  \\ 
NiD026 & 12:37:34.56 & 62:20:16.6 &  6.54 & $26.32 \pm 0.24$ & $>1.67$ &  0.62 &  \\ 
NiD027 & 12:35:48.66 & 62:12:13.3 &  6.54 & $25.67 \pm 0.28$ & $>1.72$ &  0.71 &  \\ 
NiD028 & 12:36:14.37 & 62:16:17.4 &  6.53 & $26.01 \pm 0.26$ & $>1.21$ &  0.21 &  \\ 
NiD029 & 12:37:32.67 & 62:14:16.6 &  6.47 & $26.28 \pm 0.22$ & $>1.83$ &  0.47 &  \\ 
NiD030 & 12:35:57.59 & 62:12:09.2 &  6.45 & $25.47 \pm 0.21$ & $>1.68$ &  0.50 &  \\ 
NiD031 & 12:36:15.08 & 62:16:34.4 &  6.41 & $26.33 \pm 0.26$ & $>1.48$ &  0.50 &  \\ 
SiD025 & 03:32:36.47 & 27:46:41.5 &  6.40 & $25.76 \pm 0.24$ & $>1.71$ &  0.42 & ISAAC \\ 
NiD032 & 12:37:10.96 & 62:19:48.0 &  6.39 & $25.82 \pm 0.27$ & $1.62 \pm 0.50$ &  0.60 &  \\ 
NiD033 & 12:37:11.40 & 62:22:17.0 &  6.32 & $26.02 \pm 0.24$ & $1.33 \pm 0.43$ &  0.42 &  \\ 
NiD034 & 12:36:45.51 & 62:18:32.7 &  6.30 & $25.86 \pm 0.22$ & $>1.93$ &  0.30 &  \\ 
NiD035 & 12:36:28.03 & 62:13:04.8 &  6.29 & $25.33 \pm 0.25$ & $>2.00$ &  0.50 &  \\ 
SiD026 & 03:32:17.25 & 27:46:46.0 &  6.27 & $26.30 \pm 0.20$ & $>1.59$ &  0.22 &  \\ 
NiD037 & 12:36:50.78 & 62:20:17.3 &  6.27 & $25.87 \pm 0.24$ & $>1.66$ &  0.52 &  \\ 
NiD036 & 12:37:29.93 & 62:12:15.4 &  6.27 & $26.08 \pm 0.25$ & $>1.50$ &  0.26 &  \\ 
SiD027 & 03:32:52.52 & 27:51:44.6 &  6.26 & $26.03 \pm 0.24$ & $>1.68$ &  0.57 &  \\ 
NiD038 & 12:36:27.56 & 62:13:28.3 &  6.25 & $26.24 \pm 0.21$ & $>1.59$ &  0.45 &  \\ 
NiD039 & 12:36:26.93 & 62:17:01.2 &  6.24 & $25.96 \pm 0.23$ & $1.54 \pm 0.50$ &  0.41 &  \\ 
SiD028 & 03:32:16.55 & 27:41:03.3 &  6.21 & $25.93 \pm 0.22$ & $1.36 \pm 0.45$ &  0.60 &  \\ 
NiD040 & 12:36:33.20 & 62:09:23.4 &  6.20 & $25.21 \pm 0.19$ & $>1.44$ &  0.70 &  \\ 
NiD041 & 12:36:49.93 & 62:08:02.9 &  6.18 & $26.86 \pm 0.31$ & $>1.47$ &  0.16 &  \\ 
SiD029 & 03:32:19.19 & 27:55:37.9 &  6.17 & $25.13 \pm 0.21$ & $>1.51$ &  0.74 &  \\ 
SiD030 & 03:32:29.33 & 27:40:14.4 &  6.16 & $26.59 \pm 0.21$ & $>1.70$ &  0.17 &  \\ 
SiD031 & 03:32:56.37 & 27:53:20.9 &  6.15 & $25.53 \pm 0.23$ & $>1.98$ &  0.53 &  \\ 
NiD042 & 12:36:31.98 & 62:08:26.3 &  6.12 & $26.65 \pm 0.23$ & $>1.76$ &  0.24 &  \\ 
NiD043 & 12:37:43.01 & 62:20:02.2 &  6.09 & $25.62 \pm 0.23$ & $>1.65$ &  0.57 &  \\ 
NiD044 & 12:36:48.50 & 62:10:47.3 &  6.08 & $26.28 \pm 0.28$ & $>1.61$ &  0.49 &  \\ 
NiD045 & 12:36:31.16 & 62:13:34.0 &  6.07 & $25.39 \pm 0.30$ & $>1.84$ &  0.70 &  \\ 
NiD046 & 12:35:48.97 & 62:12:25.1 &  6.07 & $25.97 \pm 0.25$ & $>1.59$ &  0.27 &  \\ 
NiD047 & 12:36:26.22 & 62:11:47.8 &  6.05 & $25.91 \pm 0.21$ & $>1.78$ &  0.52 &  \\ 
SiD032 & 03:32:42.08 & 27:41:37.2 &  6.03 & $26.23 \pm 0.29$ & $>1.51$ &  0.54 &  \\ 
SiD033 & 03:32:39.45 & 27:40:26.4 &  6.02 & $26.47 \pm 0.31$ & $>1.62$ &  0.39 &  \\ 
NiD048 & 12:36:28.26 & 62:08:19.9 &  5.99 & $26.40 \pm 0.24$ & $>1.45$ &  0.33 &  \\ 
SiD035 & 03:32:14.90 & 27:41:02.7 &  5.98 & $26.36 \pm 0.30$ & $>1.72$ &  0.20 &  \\ 
SiD034 & 03:32:27.39 & 27:47:28.3 &  5.98 & $26.11 \pm 0.28$ & $>1.65$ &  0.49 & ISAAC \\ 
NiD049 & 12:36:19.17 & 62:12:19.6 &  5.98 & $25.93 \pm 0.31$ & $>1.73$ &  0.42 &  \\ 
NiD050 & 12:37:25.65 & 62:17:43.4 &  5.97 & $26.14 \pm 0.26$ & $>1.61$ &  0.50 &  \\ 
NiD052 & 12:37:37.21 & 62:19:35.8 &  5.95 & $26.76 \pm 0.35$ & $>1.69$ &  0.27 &  \\ 
NiD051 & 12:37:40.42 & 62:13:29.5 &  5.95 & $25.58 \pm 0.26$ & $>2.06$ &  0.55 &  \\ 
NiD053 & 12:37:33.19 & 62:16:42.0 &  5.94 & $25.95 \pm 0.22$ & $1.43 \pm 0.52$ &  0.53 &  \\ 
NiD054 & 12:37:10.40 & 62:11:22.1 &  5.94 & $26.95 \pm 0.26$ & $>1.57$ &  0.16 &  \\ 
SiD036 & 03:32:14.75 & 27:45:41.6 &  5.92 & $26.40 \pm 0.26$ & $>1.62$ &  0.24 & ISAAC \\ 
NiD055 & 12:36:19.49 & 62:15:43.3 &  5.92 & $26.05 \pm 0.25$ & $>1.58$ &  0.27 &  \\ 
SiD038 & 03:32:44.70 & 27:50:02.2 &  5.90 & $26.52 \pm 0.29$ & $>1.76$ &  0.29 &  \\ 
SiD037 & 03:32:22.52 & 27:56:27.5 &  5.90 & $26.25 \pm 0.26$ & $>1.68$ &  0.45 &  \\ 
NiD056 & 12:37:34.22 & 62:15:23.2 &  5.89 & $26.00 \pm 0.28$ & $1.63 \pm 0.56$ &  0.56 &  \\ 
NiD057 & 12:35:47.07 & 62:12:18.7 &  5.88 & $26.50 \pm 0.31$ & $>1.58$ &  0.21 &  \\ 
SiD039 & 03:32:21.62 & 27:50:04.4 &  5.87 & $25.39 \pm 0.25$ & $>1.61$ &  0.69 & ISAAC \\ 
NiD058 & 12:35:53.25 & 62:10:45.3 &  5.87 & $24.99 \pm 0.16$ & $1.31 \pm 0.46$ &  0.62 &  \\ 
NiD059 & 12:36:28.86 & 62:12:22.6 &  5.86 & $26.12 \pm 0.21$ & $>1.99$ &  0.22 &  \\ 
NiD060 & 12:36:10.31 & 62:10:42.6 &  5.85 & $25.93 \pm 0.22$ & $>1.59$ &  0.58 &  \\ 
NiD061 & 12:36:25.69 & 62:15:09.5 &  5.82 & $26.32 \pm 0.26$ & $>1.68$ &  0.18 &  \\ 
NiD062 & 12:37:43.82 & 62:17:26.5 &  5.80 & $26.16 \pm 0.25$ & $>1.60$ &  0.47 &  \\ 
NiD064 & 12:37:17.86 & 62:18:20.8 &  5.79 & $26.17 \pm 0.25$ & $>1.49$ &  0.51 &  \\ 
NiD063 & 12:36:37.53 & 62:12:36.3 &  5.79 & $26.30 \pm 0.25$ & $1.63 \pm 0.38$ &  0.43 &  \\ 
NiD065 & 12:36:58.46 & 62:21:22.4 &  5.78 & $26.40 \pm 0.34$ & $>1.64$ &  0.47 &  \\ 
NiD066 & 12:37:16.14 & 62:13:01.0 &  5.78 & $25.98 \pm 0.25$ & $>1.59$ &  0.45 &  \\ 
NiD067 & 12:36:22.71 & 62:08:37.4 &  5.77 & $25.76 \pm 0.24$ & $>1.62$ &  0.39 &  \\ 
NiD068 & 12:36:48.54 & 62:18:50.7 &  5.76 & $26.09 \pm 0.30$ & $>1.54$ &  0.49 &  \\ 
NiD070 & 12:36:42.10 & 62:09:02.4 &  5.76 & $26.52 \pm 0.30$ & $>1.27$ &  0.51 &  \\ 
NiD069 & 12:35:46.99 & 62:12:28.6 &  5.76 & $25.82 \pm 0.23$ & $>1.55$ &  0.45 &  \\ 
NiD071 & 12:37:01.30 & 62:21:28.2 &  5.75 & $26.36 \pm 0.23$ & $1.42 \pm 0.50$ &  0.38 &  \\ 
NiD072 & 12:37:35.60 & 62:14:45.2 &  5.75 & $25.82 \pm 0.24$ & $>1.71$ &  0.42 &  \\ 
NiD074 & 12:35:56.18 & 62:11:45.7 &  5.74 & $26.18 \pm 0.23$ & $1.47 \pm 0.53$ &  0.48 &  \\ 
NiD073 & 12:36:10.32 & 62:08:10.9 &  5.74 & $26.86 \pm 0.22$ & $1.31 \pm 0.47$ &  0.23 &  \\ 
NiD076 & 12:37:33.97 & 62:19:30.5 &  5.73 & $26.36 \pm 0.24$ & $>1.62$ &  0.30 &  \\ 
NiD075 & 12:37:28.86 & 62:13:21.4 &  5.73 & $25.99 \pm 0.26$ & $>1.77$ &  0.23 &  \\ 
NiD077 & 12:37:38.49 & 62:19:50.9 &  5.71 & $25.96 \pm 0.27$ & $>1.56$ &  0.57 &  \\ 
NiD078 & 12:36:28.30 & 62:13:19.7 &  5.70 & $25.40 \pm 0.22$ & $>1.87$ &  0.35 &  \\ 
SiD040 & 03:32:17.95 & 27:48:16.3 &  5.68 & $25.90 \pm 0.23$ & $>1.72$ &  0.28 & ISAAC \\ 
SiD041 & 03:32:44.37 & 27:54:19.1 &  5.68 & $25.78 \pm 0.23$ & $>1.61$ &  0.26 &  \\ 
NiD079 & 12:36:36.34 & 62:16:49.4 &  5.68 & $26.25 \pm 0.28$ & $>1.67$ &  0.14 &  \\ 
NiD082 & 12:37:37.94 & 62:19:33.5 &  5.67 & $26.03 \pm 0.18$ & $>1.58$ &  0.60 &  \\ 
NiD081 & 12:36:09.46 & 62:15:12.6 &  5.67 & $26.00 \pm 0.27$ & $>1.50$ &  0.54 &  \\ 
NiD080 & 12:36:21.37 & 62:09:23.4 &  5.67 & $26.51 \pm 0.31$ & $>1.65$ &  0.42 &  \\ 
SiD042 & 03:32:22.08 & 27:42:35.9 &  5.66 & $26.32 \pm 0.27$ & $>1.69$ &  0.34 &  \\ 
NiD083 & 12:37:33.12 & 62:18:04.6 &  5.66 & $26.43 \pm 0.24$ & $>1.55$ &  0.46 &  \\ 
NiD084 & 12:35:48.44 & 62:13:04.6 &  5.66 & $26.34 \pm 0.20$ & $>1.22$ &  0.49 &  \\ 
SiD043 & 03:32:50.79 & 27:47:46.7 &  5.65 & $25.97 \pm 0.24$ & $>1.65$ &  0.20 &  \\ 
NiD085 & 12:37:19.75 & 62:16:03.1 &  5.63 & $25.93 \pm 0.25$ & $>1.63$ &  0.38 &  \\ 
SiD044 & 03:32:17.78 & 27:48:13.5 &  5.62 & $26.06 \pm 0.26$ & $1.63 \pm 0.53$ &  0.40 & ISAAC \\ 
NiD087 & 12:37:34.18 & 62:20:55.3 &  5.60 & $25.63 \pm 0.23$ & $>1.48$ &  0.56 &  \\ 
NiD086 & 12:36:46.21 & 62:18:41.7 &  5.60 & $26.36 \pm 0.34$ & $>1.53$ &  0.44 &  \\ 
SiD045 & 03:32:16.66 & 27:47:40.0 &  5.59 & $26.05 \pm 0.22$ & $>1.46$ &  0.52 & ISAAC \\ 
NiD089 & 12:36:49.61 & 62:10:39.3 &  5.58 & $26.01 \pm 0.25$ & $>1.40$ &  0.48 &  \\ 
NiD088 & 12:35:54.17 & 62:13:50.4 &  5.58 & $26.68 \pm 0.26$ & $>1.16$ &  0.36 &  \\ 
SiD046 & 03:32:04.52 & 27:45:55.3 &  5.55 & $26.17 \pm 0.33$ & $>1.57$ &  0.53 & ISAAC \\ 
NiD090 & 12:37:40.76 & 62:19:46.3 &  5.55 & $25.13 \pm 0.20$ & $>1.55$ &  0.68 &  \\ 
SiD047 & 03:32:44.47 & 27:48:21.2 &  5.54 & $25.56 \pm 0.21$ & $>1.63$ &  0.32 & ISAAC \\ 
NiD091 & 12:36:16.97 & 62:12:32.5 &  5.54 & $26.34 \pm 0.29$ & $>1.53$ &  0.19 &  \\ 
SiD048 & 03:32:27.89 & 27:43:15.8 &  5.53 & $26.57 \pm 0.23$ & $>1.52$ &  0.30 &  \\ 
NiD092 & 12:37:09.98 & 62:12:26.9 &  5.53 & $26.24 \pm 0.30$ & $>1.32$ &  0.44 &  \\ 
SiD049 & 03:32:20.50 & 27:54:34.6 &  5.52 & $26.09 \pm 0.25$ & $>1.34$ &  0.47 &  \\ 
NiD093 & 12:36:11.20 & 62:11:07.5 &  5.52 & $25.91 \pm 0.27$ & $>1.49$ &  0.55 &  \\ 
SiD050 & 03:32:53.84 & 27:51:49.2 &  5.51 & $26.12 \pm 0.26$ & $>1.44$ &  0.28 &  \\ 
NiD094 & 12:36:57.72 & 62:12:23.9 &  5.50 & $26.30 \pm 0.24$ & $>1.54$ &  0.19 &  \\ 
SiD051 & 03:32:28.34 & 27:43:15.9 &  5.49 & $27.30 \pm 0.39$ & $>1.58$ &  0.17 &  \\ 
NiD095 & 12:36:57.87 & 62:19:30.7 &  5.49 & $26.44 \pm 0.29$ & $>1.54$ &  0.34 &  \\ 
SiD052 & 03:32:05.46 & 27:46:44.2 &  5.48 & $26.40 \pm 0.26$ & $>1.41$ &  0.44 &  \\ 
NiD096 & 12:36:56.99 & 62:14:05.4 &  5.48 & $26.90 \pm 0.29$ & $>1.63$ &  0.23 &  \\ 
SiD055 & 03:32:34.75 & 27:40:35.2 &  5.47 & $26.65 \pm 0.23$ & $>1.58$ &  0.29 &  \\ 
SiD054 & 03:32:16.64 & 27:47:39.6 &  5.47 & $26.12 \pm 0.24$ & $>1.45$ &  0.46 & ISAAC \\ 
SiD053 & 03:32:35.53 & 27:53:37.2 &  5.47 & $25.72 \pm 0.26$ & $>1.55$ &  0.49 &  \\ 
SiD056 & 03:32:43.49 & 27:45:29.2 &  5.46 & $26.40 \pm 0.29$ & $>1.46$ &  0.56 &  \\ 
NiD097 & 12:37:29.90 & 62:14:08.9 &  5.46 & $25.93 \pm 0.28$ & $>1.68$ &  0.49 &  \\ 
NiD099 & 12:36:45.40 & 62:18:02.7 &  5.45 & $25.70 \pm 0.22$ & $>1.67$ &  0.42 &  \\ 
NiD098 & 12:37:07.87 & 62:09:16.9 &  5.45 & $26.36 \pm 0.28$ & $>1.28$ &  0.34 &  \\ 
NiD100 & 12:38:00.87 & 62:16:11.6 &  5.44 & $26.38 \pm 0.33$ & $>1.59$ &  0.43 &  \\ 
NiD102 & 12:36:57.20 & 62:10:24.7 &  5.43 & $26.29 \pm 0.29$ & $>1.48$ &  0.43 &  \\ 
NiD101 & 12:36:00.10 & 62:13:23.6 &  5.43 & $26.32 \pm 0.29$ & $>1.54$ &  0.27 &  \\ 
SiD057 & 03:32:37.96 & 27:42:07.6 &  5.42 & $25.91 \pm 0.26$ & $>1.58$ &  0.24 &  \\ 
SiD058 & 03:32:54.86 & 27:48:39.9 &  5.42 & $25.68 \pm 0.29$ & $>1.39$ &  0.36 &  \\ 
NiD103 & 12:37:31.68 & 62:20:18.7 &  5.42 & $25.99 \pm 0.29$ & $>1.46$ &  0.52 &  \\ 
SiD059 & 03:32:40.70 & 27:53:26.0 &  5.41 & $26.36 \pm 0.29$ & $>1.61$ &  0.40 &  \\ 
NiD104 & 12:36:58.84 & 62:10:34.5 &  5.41 & $26.60 \pm 0.22$ & $>1.49$ &  0.35 &  \\ 
NiD105 & 12:36:22.73 & 62:14:22.0 &  5.40 & $26.74 \pm 0.29$ & $1.33 \pm 0.52$ &  0.26 &  \\ 
NiD106 & 12:35:45.44 & 62:12:30.6 &  5.39 & $26.50 \pm 0.30$ & $>1.52$ &  0.37 &  \\ 
SiD061 & 03:32:19.90 & 27:52:06.1 &  5.38 & $26.88 \pm 0.26$ & $1.54 \pm 0.56$ &  0.26 &  \\ 
SiD060 & 03:32:36.67 & 27:54:21.0 &  5.38 & $26.65 \pm 0.25$ & $>1.64$ &  0.41 &  \\ 
NiD107 & 12:36:12.10 & 62:14:38.1 &  5.38 & $25.84 \pm 0.23$ & $>1.53$ &  0.39 &  \\ 
NiD108 & 12:35:39.26 & 62:12:29.3 &  5.38 & $26.14 \pm 0.32$ & $>1.12$ &  0.36 &  \\ 
NiD109 & 12:37:09.14 & 62:22:50.6 &  5.37 & $25.69 \pm 0.29$ & $>1.29$ &  0.52 &  \\ 
SiD062 & 03:32:35.98 & 27:46:05.1 &  5.36 & $26.70 \pm 0.41$ & $1.40 \pm 0.54$ &  0.29 & ISAAC \\ 
NiD111 & 12:36:20.91 & 62:16:50.8 &  5.36 & $26.87 \pm 0.40$ & $>1.63$ &  0.39 &  \\ 
NiD110 & 12:35:47.37 & 62:11:33.2 &  5.36 & $26.82 \pm 0.27$ & $>1.47$ &  0.37 &  \\ 
SiD063 & 03:32:22.39 & 27:48:04.4 &  5.35 & $26.64 \pm 0.31$ & $>1.46$ &  0.44 & ISAAC \\ 
NiD113 & 12:36:29.26 & 62:16:31.6 &  5.35 & $25.75 \pm 0.21$ & $>1.54$ &  0.38 &  \\ 
NiD112 & 12:36:22.03 & 62:15:13.8 &  5.35 & $25.83 \pm 0.25$ & $>1.45$ &  0.48 &  \\ 
NiD114 & 12:36:47.23 & 62:09:55.5 &  5.35 & $26.35 \pm 0.29$ & $>1.43$ &  0.42 &  \\ 
NiD116 & 12:36:44.51 & 62:10:28.3 &  5.34 & $26.05 \pm 0.26$ & $>1.36$ &  0.49 &  \\ 
NiD115 & 12:35:57.93 & 62:13:51.6 &  5.34 & $26.17 \pm 0.28$ & $>1.14$ &  0.28 &  \\ 
NiD118 & 12:37:16.16 & 62:18:14.9 &  5.33 & $26.24 \pm 0.26$ & $>1.52$ &  0.56 &  \\ 
NiD117 & 12:37:09.62 & 62:18:14.7 &  5.33 & $26.09 \pm 0.27$ & $>1.45$ &  0.60 &  \\ 
NiD119 & 12:36:12.62 & 62:13:48.0 &  5.32 & $26.37 \pm 0.30$ & $>1.28$ &  0.34 &  \\ 
SiD064 & 03:32:24.80 & 27:47:58.8 &  5.31 & $26.53 \pm 0.29$ & $>1.54$ &  0.35 & ISAAC \\ 
SiD065 & 03:32:40.82 & 27:47:43.1 &  5.31 & $26.32 \pm 0.23$ & $>1.64$ &  0.51 & ISAAC \\ 
SiD066 & 03:32:44.12 & 27:43:18.4 &  5.30 & $25.91 \pm 0.27$ & $>1.49$ &  0.48 &  \\ 
NiD120 & 12:36:28.07 & 62:13:19.8 &  5.30 & $25.89 \pm 0.29$ & $1.50 \pm 0.52$ &  0.37 &  \\ 
NiD121 & 12:37:15.05 & 62:18:17.8 &  5.28 & $25.60 \pm 0.23$ & $>1.46$ &  0.54 &  \\ 
NiD123 & 12:37:14.84 & 62:20:15.0 &  5.27 & $26.17 \pm 0.23$ & $1.40 \pm 0.50$ &  0.49 &  \\ 
NiD122 & 12:36:04.56 & 62:09:24.7 &  5.27 & $26.05 \pm 0.27$ & $>1.43$ &  0.49 &  \\ 
SiD067 & 03:32:19.46 & 27:51:59.2 &  5.26 & $26.34 \pm 0.32$ & $1.34 \pm 0.50$ &  0.43 &  \\ 
SiD068 & 03:32:53.20 & 27:49:44.3 &  5.26 & $26.17 \pm 0.27$ & $>1.53$ &  0.31 &  \\ 
NiD125 & 12:37:50.65 & 62:17:22.4 &  5.26 & $26.43 \pm 0.31$ & $>1.63$ &  0.40 &  \\ 
NiD124 & 12:36:37.49 & 62:16:57.5 &  5.26 & $26.57 \pm 0.32$ & $>1.60$ &  0.38 &  \\ 
NiD126 & 12:36:17.37 & 62:16:17.7 &  5.26 & $26.43 \pm 0.30$ & $>1.41$ &  0.47 &  \\ 
SiD069 & 03:32:13.06 & 27:51:33.6 &  5.25 & $26.41 \pm 0.20$ & $>1.33$ &  0.26 &  \\ 
SiD070 & 03:32:39.19 & 27:54:13.8 &  5.25 & $26.55 \pm 0.33$ & $>1.61$ &  0.43 &  \\ 
NiD128 & 12:36:57.56 & 62:09:08.5 &  5.25 & $25.60 \pm 0.24$ & $>1.38$ &  0.54 &  \\ 
NiD127 & 12:36:12.41 & 62:14:49.5 &  5.25 & $25.75 \pm 0.22$ & $>1.52$ &  0.45 &  \\ 
SiD071 & 03:32:14.73 & 27:47:58.7 &  5.23 & $26.12 \pm 0.33$ & $1.68 \pm 0.47$ &  0.21 & ISAAC \\ 
SiD072 & 03:32:47.69 & 27:46:45.1 &  5.23 & $25.13 \pm 0.18$ & $>1.54$ &  0.54 &  \\ 
NiD129 & 12:37:36.60 & 62:14:09.7 &  5.22 & $26.46 \pm 0.25$ & $>1.62$ &  0.45 &  \\ 
SiD073 & 03:32:19.05 & 27:42:44.2 &  5.21 & $25.66 \pm 0.23$ & $>1.54$ &  0.51 &  \\ 
NiD130 & 12:37:31.00 & 62:19:49.0 &  5.21 & $26.37 \pm 0.30$ & $>1.47$ &  0.42 &  \\ 
NiD132 & 12:37:17.48 & 62:17:46.3 &  5.21 & $25.95 \pm 0.26$ & $>1.49$ &  0.56 &  \\ 
NiD131 & 12:36:48.78 & 62:19:39.0 &  5.21 & $26.49 \pm 0.21$ & $>1.21$ &  0.45 &  \\ 
SiD074 & 03:32:34.82 & 27:51:33.1 &  5.20 & $26.32 \pm 0.30$ & $>1.69$ &  0.26 &  \\ 
NiD133 & 12:37:42.08 & 62:15:04.4 &  5.20 & $25.98 \pm 0.26$ & $>1.53$ &  0.28 &  \\ 
NiD134 & 12:36:24.96 & 62:10:56.0 &  5.20 & $26.41 \pm 0.26$ & $>1.39$ &  0.23 &  \\ 
NiD136 & 12:36:51.49 & 62:20:10.8 &  5.18 & $26.88 \pm 0.28$ & $>1.47$ &  0.20 &  \\ 
NiD135 & 12:36:37.86 & 62:14:26.8 &  5.18 & $26.38 \pm 0.20$ & $>1.44$ &  0.47 &  \\ 
SiD075 & 03:32:23.88 & 27:52:04.4 &  5.17 & $26.25 \pm 0.20$ & $>1.61$ &  0.43 &  \\ 
NiD137 & 12:36:27.03 & 62:11:25.9 &  5.16 & $25.81 \pm 0.30$ & $>1.79$ &  0.36 &  \\ 
SiD076 & 03:32:19.90 & 27:47:53.2 &  5.15 & $25.97 \pm 0.28$ & $>1.41$ &  0.44 & ISAAC \\ 
SiD077 & 03:32:21.60 & 27:44:23.0 &  5.14 & $26.60 \pm 0.30$ & $>1.51$ &  0.41 & ISAAC \\ 
SiD078 & 03:32:18.54 & 27:52:59.9 &  5.14 & $26.03 \pm 0.29$ & $>1.59$ &  0.51 &  \\ 
NiD139 & 12:37:39.99 & 62:20:08.4 &  5.14 & $26.93 \pm 0.24$ & $>1.55$ &  0.34 &  \\ 
NiD138 & 12:37:41.69 & 62:19:29.4 &  5.14 & $26.22 \pm 0.23$ & $>1.47$ &  0.44 &  \\ 
NiD140 & 12:37:15.31 & 62:15:35.7 &  5.14 & $26.18 \pm 0.29$ & $1.47 \pm 0.56$ &  0.46 &  \\ 
SiD079 & 03:32:29.41 & 27:43:49.4 &  5.13 & $26.21 \pm 0.28$ & $>1.49$ &  0.28 &  \\ 
NiD141 & 12:36:29.43 & 62:16:44.4 &  5.13 & $26.83 \pm 0.30$ & $>1.55$ &  0.29 &  \\ 
NiD144 & 12:36:41.38 & 62:17:01.9 &  5.11 & $26.76 \pm 0.28$ & $>1.58$ &  0.29 &  \\ 
NiD142 & 12:37:01.34 & 62:11:39.7 &  5.11 & $25.91 \pm 0.26$ & $>1.43$ &  0.57 &  \\ 
NiD143 & 12:36:49.09 & 62:09:12.6 &  5.11 & $26.46 \pm 0.34$ & $>1.32$ &  0.46 &  \\ 
SiD080 & 03:32:18.29 & 27:48:55.6 &  5.10 & $26.63 \pm 0.22$ & $1.43 \pm 0.58$ &  0.34 & ISAAC \\ 
SiD082 & 03:32:16.26 & 27:44:19.7 &  5.09 & $26.68 \pm 0.27$ & $>1.37$ &  0.34 & ISAAC \\ 
SiD083 & 03:32:05.13 & 27:46:40.0 &  5.09 & $26.44 \pm 0.33$ & $>1.51$ &  0.30 &  \\ 
SiD081 & 03:32:52.36 & 27:48:53.0 &  5.09 & $26.19 \pm 0.26$ & $>1.54$ &  0.47 &  \\ 
SiD085 & 03:32:49.63 & 27:49:11.1 &  5.08 & $26.66 \pm 0.34$ & $>1.52$ &  0.23 &  \\ 
SiD084 & 03:32:23.37 & 27:51:55.7 &  5.08 & $25.96 \pm 0.25$ & $>1.54$ &  0.27 &  \\ 
NiD145 & 12:36:50.61 & 62:10:52.7 &  5.08 & $26.58 \pm 0.21$ & $>1.06$ &  0.44 &  \\ 
NiD146 & 12:36:44.70 & 62:10:03.1 &  5.08 & $26.84 \pm 0.27$ & $>1.45$ &  0.40 &  \\ 
SiD086 & 03:32:41.17 & 27:49:47.8 &  5.06 & $26.00 \pm 0.24$ & $>1.42$ &  0.34 &  \\ 
SiD088 & 03:32:39.97 & 27:41:50.0 &  5.05 & $25.89 \pm 0.26$ & $>1.28$ &  0.47 &  \\ 
SiD087 & 03:32:42.16 & 27:54:38.8 &  5.05 & $25.86 \pm 0.27$ & $>1.53$ &  0.52 &  \\ 
NiD147 & 12:37:13.52 & 62:16:20.0 &  5.05 & $26.55 \pm 0.33$ & $>1.59$ &  0.34 &  \\ 
SiD089 & 03:32:29.02 & 27:42:08.0 &  5.04 & $26.94 \pm 0.24$ & $>1.52$ &  0.22 &  \\ 
NiD148 & 12:36:34.17 & 62:16:47.1 &  5.04 & $26.71 \pm 0.25$ & $>1.55$ &  0.42 &  \\ 
SiD092 & 03:32:09.91 & 27:43:36.3 &  5.03 & $25.41 \pm 0.21$ & $>1.69$ &  0.50 & ISAAC \\ 
SiD090 & 03:32:33.78 & 27:48:07.6 &  5.03 & $26.29 \pm 0.30$ & $>1.49$ &  0.39 & ISAAC \\ 
SiD091 & 03:32:49.83 & 27:48:38.3 &  5.03 & $26.40 \pm 0.26$ & $>1.49$ &  0.40 &  \\ 
NiD149 & 12:37:30.73 & 62:19:44.7 &  5.03 & $26.05 \pm 0.27$ & $>1.42$ &  0.41 &  \\ 
SiD095 & 03:32:07.57 & 27:41:30.3 &  5.02 & $26.12 \pm 0.31$ & $>1.74$ &  0.39 &  \\ 
SiD093 & 03:32:21.35 & 27:50:30.6 &  5.02 & $26.46 \pm 0.29$ & $1.39 \pm 0.51$ &  0.33 & ISAAC \\ 
SiD094 & 03:32:21.75 & 27:50:52.0 &  5.02 & $26.32 \pm 0.31$ & $>1.44$ &  0.50 & ISAAC \\ 
SiD096 & 03:32:20.72 & 27:44:35.3 &  5.01 & $26.62 \pm 0.24$ & $>1.52$ &  0.42 & ISAAC \\ 
NiD153 & 12:36:55.43 & 62:20:50.5 &  5.01 & $26.43 \pm 0.30$ & $>1.38$ &  0.36 &  \\ 
NiD151 & 12:37:08.17 & 62:09:42.7 &  5.01 & $26.35 \pm 0.32$ & $>1.10$ &  0.41 &  \\ 
NiD154 & 12:36:29.21 & 62:13:35.6 &  5.01 & $26.80 \pm 0.33$ & $>1.41$ &  0.13 &  \\ 
NiD152 & 12:36:27.41 & 62:12:05.3 &  5.01 & $25.89 \pm 0.29$ & $>1.75$ &  0.25 &  \\ 
NiD150 & 12:36:25.29 & 62:11:41.6 &  5.01 & $26.22 \pm 0.30$ & $>1.67$ &  0.39 &  \\ 
NiD155 & 12:36:24.54 & 62:15:35.8 &  5.00 & $26.93 \pm 0.28$ & $>1.33$ &  0.41 &  \\ 

\enddata
\tablenotetext{a}{The robust sample is that with $S/N(\acsz) > 10$;  objects below
the line should be regarded with caution, and may include spurious contaminants
(see text).}
\tablenotetext{b}{Color limits are reported at $2\sigma$}

\end{deluxetable}


\end{document}